\documentclass[12pt]{article}

\usepackage{bbm,latexsym,graphicx}

\newcommand{\be}{\begin{equation}}
\newcommand{\ee}{\end{equation}}
\newcommand{\ba}{\begin{eqnarray}}
\newcommand{\ea}{\end{eqnarray}}
\newcommand{\no}{\nonumber\\}

\newcommand{\lesssim}{ \ \mbox{\raisebox{-3pt}{$\stackrel%
{\displaystyle <}{\sim}$}} \ }
\newcommand{\gtrsim}{\:\mbox{\raisebox{-3pt}{$\stackrel%
{\displaystyle >}{\sim}$}}\:}

\newcommand{\mnu}{\mathcal{M}_\nu}

\begin{document}

\title{\normalsize \hfill UWThPh-2005-5 \\[1cm]
\LARGE $S_3 \times \mathbbm{Z}_2$ model
for neutrino mass matrices}
\author{Walter Grimus\thanks{E-mail:
walter.grimus@univie.ac.at} \\
\setcounter{footnote}{6}
\small Institut f\"ur Theoretische Physik, Universit\"at Wien \\
\small Boltzmanngasse 5, A--1090 Wien, Austria
\\*[3.6mm]
Lu\'{\i}s Lavoura\thanks{E-mail:
balio@cftp.ist.utl.pt} \\
\small Universidade T\'ecnica de Lisboa
and Centro de F\'\i sica Te\'orica de Part\'\i culas \\
\small Instituto Superior T\'ecnico, 1049-001 Lisboa, Portugal
\\*[4.6mm] }

\date{2 August 2005}

\maketitle

\begin{abstract}
We propose a model for lepton mass matrices
based on the seesaw mechanism,
a complex scalar gauge singlet
and a horizontal symmetry
$S_3 \times \mathbbm{Z}_2$.
In a suitable weak basis,
the charged-lepton mass matrix
and the neutrino Dirac mass matrix are diagonal,
but the vacuum expectation value
of the scalar gauge singlet
renders the Majorana mass matrix
of the right-handed neutrinos non-diagonal,
thereby generating lepton mixing.
When the symmetry $S_3$ is not broken in the scalar potential,
the effective light-neutrino Majorana mass matrix
enjoys $\mu$--$\tau$ interchange symmetry,
thus predicting maximal atmospheric neutrino mixing
together with $U_{e3} = 0$.
A partial and less predictive
form of $\mu$--$\tau$ interchange symmetry
is obtained when the symmetry $S_3$ is softly broken
in the scalar potential.
Enlarging the symmetry group $S_3 \times \mathbbm{Z}_2$
by an additional discrete electron-number symmetry 
$\mathbbm{Z}_2^{(e)}$,
a more predicitive model is obtained,
which is in practice indistinguishable
from a previous one based on the group $D_4$. 
\end{abstract}

\newpage

\paragraph{Introduction}
The very precise data now existing
on neutrino mass-squared differences
and on lepton mixing~\cite{tortola},
and the prospects of rapid experimental developments
in this field,
invite theorists to construct models
for the lepton mass matrices,
in an effort to exploit and to understand
the symmetries and hierarchies suggested by the data.
Among them,
most prominent are the possible maximality
of the atmospheric neutrino mixing angle $\theta_{23}$
and the smallness of the mixing-matrix element $U_{e3}$.
Together,
they suggest the existence
of a $\mu$--$\tau$ interchange symmetry
in the (effective) light-neutrino Majorana mass matrix $\mnu$,
taken in the basis where the charged-lepton mass matrix
is diagonal~\cite{early}.
Such a symmetry,
embodied in
\be
\mnu = \left( \begin{array}{ccc}
x & y & y \\ y & z & w \\ y & w & z
\end{array} \right),
\label{symmetric_mnu}
\ee
automatically leads to,
simultaneously,
$U_{e3} = 0$ and $\theta_{23} = \pi / 4$.
Various authors have dwelt
on the matrix~(\ref{symmetric_mnu}),
and on generalizations thereof,
in the past~\cite{matrix1,softD4,koide,aizawa};
we,
in particular,
have shown that it may be obtained
either from a model based
on family lepton-number symmetries~\cite{z2}
or from a model based
on the discrete eight-element group $D_4$~\cite{d4}.

We show in this letter
that the matrix~(\ref{symmetric_mnu})
may also be obtained from a model
based on the smaller discrete group $S_3$,
a group which has a long tradition
in model building~\cite{pakvasa}.
The model presented here also suggests a generalization
of the matrix~(\ref{symmetric_mnu}),
wherein
\be
\mnu^{-1} =
\left( \begin{array}{ccc}
r & s & s \\ s & p e^{i \psi} & q \\ s & q & p e^{- i \psi}
\end{array} \right).
\label{general_mnu}
\ee
This generalization,
while leading to neither $U_{e3} = 0$
nor $\theta_{23} = \pi / 4$,
seems interesting in itself.

We note that,
in~\cite{softD4},
the mass matrix~(\ref{symmetric_mnu})
has been generalized in such a way that 
$\theta_{23}$ differs from $\pi/4$,
while 
$U_{e3} = 0$ remains intact; on the other hand,
with matrix~(\ref{general_mnu})---as we
shall see later---the deviation
of $U_{e3}$ from zero is correlated with
the deviation of $\theta_{23}$ from $\pi / 4$.

\paragraph{The model}
We work in the context
of a non-supersymmetric $SU(2)_L \times U(1)$ framework.
The three left-handed lepton $SU(2)_L$ doublets
are denoted by $D_{e, \mu, \tau}$.
The three right-handed charged-lepton
$SU(2)_L$ singlets are $e_R$,
$\mu_R$ and $\tau_R$.
We further introduce three $SU(2)_L$ singlet
right-handed neutrinos $\nu_{eR}$,
$\nu_{\mu R}$ and $\nu_{\tau R}$,
in order to enable the seesaw mechanism
for suppressing the light-neutrino masses~\cite{seesaw}.
In our model there are three Higgs
$SU(2)_L$ doublets $\phi_{1,2,3}$.
In exact analogy to the $D_4$ model~\cite{d4},
we introduce a symmetry
\be
\mathbbm{Z}_2^{\rm (aux)}:
\quad
\nu_{eR},\ \nu_{\mu R},\ \nu_{\tau R},\ \phi_1,\ e_R\ \,
\mbox{change\ sign}.
\ee
Instead of \emph{two real} neutral scalar $SU(2)_L$ singlets,
as in~\cite{d4},
we use \emph{one complex} neutral scalar $SU(2)_L$ singlet,
$\chi$.
Again in exact analogy to~\cite{d4},
we define a symmetry
\be
\mathbbm{Z}_2^{\rm (tr)}:
\quad
D_\mu \leftrightarrow D_\tau,\
\mu_R \leftrightarrow \tau_R,\
\nu_{\mu R} \leftrightarrow \nu_{\tau R},\
\chi \to \chi^\ast,\
\phi_3 \to - \phi_3.
\label{Z2tr}
\ee
This $\mathbbm{Z}_2^{\rm (tr)}$
is the $\mu$--$\tau$ interchange symmetry.
The crucial difference between the $D_4$ model~\cite{d4}
and the present $S_3$ one is that,
while in the $D_4$ model there was a symmetry
$\mathbbm{Z}_2^{(\tau)}$ which,
together with $\mathbbm{Z}_2^{\rm (tr)}$,
generated a group $D_4$,
in the $S_3$ model we introduce instead a symmetry
$\mathbbm{Z}_3$ which,
together with $\mathbbm{Z}_2^{\rm (tr)}$,
generates a group $S_3$~\cite{S3}.
With $\omega \equiv \exp{\left( 2 i \pi / 3 \right)}$,
we impose
\be\
\mathbbm{Z}_3:
\quad
\begin{array}{ll}
D_\mu \to \omega D_\mu, & D_\tau \to \omega^2 D_\tau,
\\*[1mm]
\mu_R \to \omega \mu_R, & \tau_R \to \omega^2 \tau_R,
\\*[1mm]
\nu_{\mu R} \to \omega \nu_{\mu R}, & \nu_{\tau R} \to \omega^2 \nu_{\tau R},
\\*[1mm]
\chi \to \omega \chi, & \chi^\ast \to \omega^2 \chi^\ast.
\end{array}
\label{z3}
\ee
Thus,
$\left( D_\mu, D_\tau \right)$,
$\left( \mu_R, \tau_R \right)$,
$\left( \nu_{\mu R}, \nu_{\tau R} \right)$
and $\left( \chi, \chi^\ast \right)$
are doublets of $S_3$.
The Higgs $SU(2)_L$ doublet $\phi_3$
changes sign under the odd permutations of $S_3$,
but stays invariant under the cyclic permutations.

The Yukawa Lagrangian symmetric
under $S_3 \times \mathbbm{Z}_2^\mathrm{(aux)}$ is
\ba
\mathcal{L}_{\rm Y} &=&
- \left[ y_1 \bar D_e \nu_{eR}
+ y_2 \left( \bar D_\mu \nu_{\mu R} + \bar D_\tau \nu_{\tau R} \right)
\right] \tilde \phi_1
\no & &
- y_3 \bar D_e e_R \phi_1
- y_4 \left( \bar D_\mu \mu_R + \bar D_\tau \tau_R \right) \phi_2
- y_5 \left( \bar D_\mu \mu_R - \bar D_\tau \tau_R \right) \phi_3
\no & &
+ y_\chi^\ast\, \nu_{eR}^T C^{-1}
\left( \nu_{\mu R} \chi^\ast + \nu_{\tau R} \chi \right)
\no & &
+ \frac{z_\chi^\ast}{2} \left( \nu_{\mu R}^T C^{-1} \nu_{\mu R} \chi
+ \nu_{\tau R}^T C^{-1} \nu_{\tau R} \chi^\ast
\right) + {\rm H.c.},
\label{yukawas}
\ea
where $\tilde \phi_1 \equiv i \tau_2 \phi_1^\ast$.
There is also an $S_3 \times
\mathbbm{Z}_2^\mathrm{(aux)}$-invariant Majorana mass term
\be
\mathcal{L}_{\rm M} = \frac{m^\ast}{2}\,
\nu_{eR}^T C^{-1} \nu_{eR}
+ {m^\prime}^\ast \nu_{\mu R}^T C^{-1} \nu_{\tau R}
+ {\rm H.c.}
\label{LM}
\ee
The second term in the right-hand side of~(\ref{LM}) differs,
in a crucial fashion,
from the analogous term in the $D_4$ model---see
equation~(9) of~\cite{d4}.

With vacuum expectation values (VEVs)
 $\left\langle 0 \left| \phi_j^0 \right| 0 \right\rangle
= v_j$ for $j = 1, 2, 3$,
one obtains
\ba
m_e &=& \left| y_3 v_1 \right|,
\no
m_\mu &=& \left| y_4 v_2 + y_5 v_3 \right|,
\\
m_\tau &=& \left| y_4 v_2 - y_5 v_3 \right|.
\nonumber
\ea
The $\mu$--$\tau$ interchange symmetry
$\mathbbm{Z}_2^\mathrm{(tr)}$
is spontaneously broken by the VEV of $\phi_3^0$,
so that the $\mu$ and $\tau$ charged leptons
acquire different masses.
The smallness of $m_\mu$ relative to $m_\tau$
may be explained
by requiring the model to be invariant under an additional, 
softly broken symmetry~\cite{mutau}.

The neutrino Dirac mass matrix is
\be
M_D = {\rm diag} \left( a, b, b \right),\ {\rm with}\
a = y_1^\ast v_1,\ b = y_2^\ast v_1.
\label{MD}
\ee
When the singlet $\chi$ acquires a VEV
$\left\langle 0 \left| \chi \right| 0 \right\rangle = W$,
one obtains Majorana mass terms for the right-handed neutrinos:
\be
\mathcal{L}_{M_R} = - \frac{1}{2} \left(
\bar \nu_{eR},\ \bar \nu_{\mu R},\ \bar \nu_{\tau R}
\right) M_R\, C \left( \begin{array}{c}
\bar \nu_{eR}^T \\ \bar \nu_{\mu R}^T \\ \bar \nu_{\tau R}^T
\end{array} \right) + \mathrm{H.c.},
\ee
with
\be
M_R = \left( \begin{array}{ccc}
m & y_\chi W & y_\chi W^\ast \\
y_\chi W & z_\chi W^\ast & m^\prime \\
y_\chi W^\ast & m^\prime & z_\chi W
\end{array} \right).
\ee
We next perform a rephasing of the fields,
\be
\begin{array}{rclcrclcrcl}
\nu_{\mu R} &\to& e^{i \alpha} \nu_{\mu R}, & &
D_\mu &\to& e^{i \alpha} D_\mu, & &
\mu_R &\to& e^{i \alpha} \mu_R,
\\
\nu_{\tau R} &\to& e^{- i \alpha} \nu_{\tau R}, & &
D_\tau &\to& e^{- i \alpha} D_\tau, & &
\tau_R &\to& e^{- i \alpha} \tau_R,
\end{array}\ \
{\rm with}\ \alpha \equiv \arg{W},
\label{rephasing}
\ee
to obtain
\be
M_R = \left( \begin{array}{ccc}
m & y_\chi \left| W \right| & y_\chi \left| W \right| \\
y_\chi \left| W \right| & z_\chi \left| W \right| e^{- 3 i \alpha}
& m^\prime \\
y_\chi \left| W \right| & m^\prime & z_\chi \left| W \right| e^{3 i \alpha}
\end{array} \right).
\label{MR}
\ee
We see that the matrix $M_R$
has become $\mu$--$\tau$ symmetric after the rephasing,
provided $W^3$ is real
($e^{3 i \alpha} = \pm 1$).
We shall see shortly that it is indeed possible
to enforce this.
If $M_R$ is $\mu$--$\tau$ symmetric,
and since $M_D$ also enjoys the $\mu$--$\tau$
interchange symmetry,
it follows,
by applying the seesaw mechanism,\footnote{We assume that $m$,
$m^\prime$ and the VEV $W$ are all of the same
very large order of magnitude.} that
\be
\mnu = - M_D^T M_R^{-1} M_D
\ee
is $\mu$--$\tau$ symmetric,
i.e.\ it is of the form~(\ref{symmetric_mnu}).

We thus find that it is possible to produce
a neutrino mass matrix of the form~(\ref{symmetric_mnu}),
which leads to $U_{e3} = 0$
and $\theta_{23} = \pi / 4$,
out of a model with symmetry $S_3 \times \mathbbm{Z}_2^\mathrm{(aux)}$
with three Higgs $SU(2)_L$ doublets---two of which
are $S_3$-invariant while the third one
changes sign under the odd permutations of $S_3$.
The charged-lepton mass matrix is automatically diagonal,
hence there are no flavour-changing neutral currents
\emph{at tree level} in the charged-lepton sector---such
interactions appear,
though,
already at the one-loop level~\cite{loop}.

\paragraph{The scalar potential} Because of the symmetry
$S_3 \times \mathbbm{Z}_2^\mathrm{(aux)}$,
the scalar potential is
\ba
V &=& \mu_\chi \left| \chi \right|^2
+ \lambda \left| \chi \right|^4
+ \sum_{j=1}^3 \left( \phi_j^\dagger \phi_j \right) \left(
\mu_j + a_j \phi_j^\dagger \phi_j
+ b_j \left| \chi \right|^2 \right)
\no & &
+ \sum_{j < k} \left[
a_{jk} \left( \phi_j^\dagger \phi_j \right)
\left( \phi_k^\dagger \phi_k \right)
+ b_{jk} \left( \phi_j^\dagger \phi_k \right)
\left( \phi_k^\dagger \phi_j \right)
+ c_{jk} \left( \phi_j^\dagger \phi_k \right)^2
+ c_{jk}^\ast \left( \phi_k^\dagger \phi_j \right)^2
\right]
\no & &
+ m_\chi \left( \chi^3 + {\chi^\ast}^3 \right).
\label{pot}
\ea
Only the term in the last line of~(\ref{pot})
feels the phase of $\chi$.
If its coefficient $m_\chi$ is negative,
then the phase of the VEV $W$ will adjust
so that $W^3$ is real and positive,
i.e.\ $\alpha$ will be either $0$ or $\pm 2 \pi / 3$;
if $m_\chi$ is positive,
then $\alpha$ will be either $\pi$ or $\pm \pi / 3$,
in order that $W^3$ is real and negative.
In any case,
$W^3$ is real.
This is precisely what is needed in order to obtain
a $\mu$--$\tau$-symmetric $\mnu$.

The situation is modified if we allow
the symmetry $\mathbbm{Z}_3$ of~(\ref{z3})
to be softly broken
by terms of dimension one,
or one and two,
while keeping both $\mathbbm{Z}_2^\mathrm{(aux)}$
and $\mathbbm{Z}_2^\mathrm{(tr)}$ unscathed.\footnote{If
we also allow the soft breaking of~(\ref{z3})
by terms of dimension three,
then Majorana mass terms
$\nu_{eR}^T C^{-1} \left( \nu_{\mu R} + \nu_{\tau R} \right)$
and $\nu_{\mu R}^T C^{-1} \nu_{\mu R}
+ \nu_{\tau R}^T C^{-1} \nu_{\tau R}$
are also present in the Lagrangian and,
after $\chi$ gets a VEV,
the $\mu$--$\tau$ interchange symmetry is destroyed altogether.}
There are only two such terms,
namely
\be
\label{soft}
\mu_\chi^\prime \left( \chi^2 + {\chi^\ast}^2 \right)
+ M \left( \chi + \chi^\ast \right),
\ee
with real constants $\mu_\chi^\prime$,
$M$.
These terms 
get added to $V$ in~(\ref{pot}),
which does not change otherwise.
The phase of $W$ becomes arbitrary.
The matrix $M_R$ in~(\ref{MR}) does not respect
$\mu$--$\tau$ interchange symmetry any more,
rather only a partial version thereof.

If one worries about cosmological domain walls,
then one may want to eliminate from the Lagrangian
all exact discrete symmetries.
This one may do by breaking $\mathbbm{Z}_2^\mathrm{(aux)}$
and $\mathbbm{Z}_2^\mathrm{(tr)}$,
together with $\mathbbm{Z}_3$,
softly by terms of dimension two.
This amounts to the addition,
to the scalar potential~(\ref{pot}),
of all terms $\phi_j^\dagger \phi_k$ with $j \neq k$.
(When $\mathbbm{Z}_2^\mathrm{(tr)}$ is broken softly,
the terms of~(\ref{soft}) also have to be generalized to 
$\mu_\chi^\prime \chi^2 + M \chi + \mbox{H.c.}$
with complex $\mu_\chi^\prime,\ M$.) 
However,
this soft breaking only affects the values of the $v_j$
and has no influence on the lepton mass matrices.
It is thus irrelevant for the following discussion.

\paragraph{Reproducing the $D_4$ model}
In the $D_4$ model~\cite{d4}
there is an \emph{accidental} symmetry
\be
\mathbbm{Z}_2^{(e)}:
\quad
D_e,\ e_R,\ \nu_{eR},\ \chi\ \, \mbox{change\ sign}.
\label{Z2e}
\ee
In the context of the present $S_3$ model,
one may promote that symmetry to \emph{fundamental}
and impose it on the Lagrangian from the start.
It enforces $z_\chi = 0$,
hence $\left( M_R \right)_{22} = \left( M_R \right)_{33} = 0$;
since $M_R = - M_D \mnu^{-1} M_D^T$
and $M_D$ is diagonal,
this means the vanishing of the $\left( \mu, \mu \right)$
and $\left( \tau, \tau \right)$ matrix elements of $\mnu^{-1}$.
The phase $\alpha$ of $W$ is irrelevant when
$\left( M_R \right)_{22} = \left( M_R \right)_{33} = 0$,
since it may be rephased away as in~(\ref{rephasing}),
and the model is automatically $\mu$--$\tau$ symmetric.
Thus, in the $S_3$ model with the extra
$\mathbbm{Z}_2^{(e)}$ symmetry one has
\be
\mathcal{M}^{-1}_\nu
= \left( \begin{array}{ccc}
r & s & s \\ s & 0 & q \\ s & q & 0
\end{array} \right),
\label{mnus3}
\ee
i.e.\ $\left( \mnu^{-1} \right)_{\mu \mu}
= \left( \mnu^{-1} \right)_{\tau \tau} = 0$.

On the other hand, in the $D_4$ model~\cite{d4} one has 
$\left( \mnu^{-1} \right)_{\mu \tau} = 0$ but 
$\left( \mnu^{-1} \right)_{\mu \mu}
= \left( \mnu^{-1} \right)_{\tau \tau} \neq 0$.
We shall demonstrate now that the matrix~(\ref{mnus3}) is equivalent
to the mass matrix of the $D_4$ model.

In general,
$\mnu$ is diagonalized as
\be
U^T \mnu U = \mathrm{diag} \left( m_1, m_2, m_3 \right),
\label{U}
\ee
where $U$ is the lepton mixing matrix and
the $m_j$ are the (real and non-negative) neutrino masses.
Equivalently,
\be
\mnu^{-1} = U\, \mathrm{diag}
\left( m_1^{-1}, m_2^{-1}, m_3^{-1} \right) U^T.
\label{mnu-1}
\ee
The unitary $U$ is parametrized as
\ba
U &=& \mathrm{diag} \left(
e^{i \vartheta_1},
e^{i \vartheta_2},
e^{i \vartheta_3} \right)
\no & & \times
\left( \begin{array}{ccc}
c_{13} c_{12} & c_{13} s_{12} & s_{13} e^{- i \delta} \\
-c_{23} s_{12} - s_{23} s_{13} c_{12} e^{i \delta} &
c_{23} c_{12} - s_{23} s_{13} s_{12} e^{i \delta} &
s_{23} c_{13} \\
-s_{23} s_{12} + c_{23} s_{13} c_{12} e^{i \delta} &
s_{23} c_{12} + c_{23} s_{13} s_{12} e^{i \delta} &
- c_{23} c_{13}
\end{array} \right)
\no & & \times\
\mathrm{diag} \left(
e^{i \Delta / 2}, 1, e^{i \Omega / 2} \right),
\label{formU}
\ea
where $c_{ij} \equiv \cos{\theta_{ij}}$,
$s_{ij} \equiv \sin{\theta_{ij}}$
and $\theta_{12}$ is the solar mixing angle.
The phases $\vartheta_i$ are unphysical;
physical are only
the Dirac phase $\delta$
and the Majorana phases $\Delta$ and $\Omega$.
In the case of $\mu$--$\tau$ symmetric $\mathcal{M}^{-1}_\nu$,
one has---see for
instance~\cite{leptogenesis}---$\vartheta_2 = \vartheta_3$,
$\theta_{13} = 0$
and $\theta_{23} = \pi / 4$; the vanishing
of $\theta_{13}$ allows one to write the 
non-diagonal matrix on the right-hand side 
of~(\ref{formU}) as a product $U_{23} U_{12}$,
where $U_{23}$ and $U_{12}$ are responsible
for the mixing in the atmospheric and solar neutrino sector,
respectively.

Now we turn to the matrix~(\ref{mnus3}),
which we can transform according to
\be
T \mathcal{M}^{-1}_\nu T
= \left( \begin{array}{ccc}
r & s & s \\ s & q & 0 \\ s & 0 & q
\end{array} \right),
\quad \mbox{with} \quad 
T = \left( \begin{array}{ccc}
1 & 0 & 0 \\ 0 & u & u^* \\ 0 & u^* & u \end{array}
\right)
\quad \mbox{and} \quad u =
\frac{e^{i\pi/4}}{\sqrt{2}}.
\label{Tmnu}
\ee
Thus,
$T \mathcal{M}^{-1}_\nu T$
has precisely the form of the mass
matrix of the $D_4$ model,
and from~(\ref{mnu-1}) it is clear
that it can be diagonalized by a
$U$ appropriate for a $\mu$--$\tau$ symmetric matrix. Since
$\vartheta_2 = \vartheta_3$, 
in the diagonalization of $T \mathcal{M}^{-1}_\nu T$
the product $T U_{23}$ occurs. Now,
\be
T U_{23} = U_{23}\, \mbox{diag} \left( 1, 1, i \right),
\quad \mbox{where} \quad 
U_{23} = \left( \begin{array}{ccc}
1 & 0 & 0 \\ 0 & \rho & \rho \\ 0 & \rho & -\rho
\end{array} \right)
\quad \mbox{and} \quad \rho = \frac {1}{\sqrt{2}}. 
\ee
The matrix $U_{12}$ commutes
with $\mbox{diag} \left( 1, 1, i \right)$.
Therefore, 
the difference between the $D_4$ model
and the $S_3 \times \mathbbm{Z}_2^\mathrm{(aux)}
\times \mathbbm{Z}_2^\mathrm{(e)}$ model
amounts to the modification $\Omega \to \Omega + \pi$.
Since $\Omega$ is a free parameter,  
the two models are in practice equivalent.

Just as in the $D_4$ model~\cite{d4},
the zero in the matrix~(\ref{mnus3}),
or---equivalently---the zero in 
$T \mnu^{-1} T$ of~(\ref{Tmnu}),
leads to
\be
\frac{s^2_{12}\, e^{i \Delta}}{m_1}
+ \frac{c^2_{12}}{m_2}
+ \frac{e^{i \Omega}}{m_3} = 0.
\label{s3rel}
\ee
As we have shown in~\cite{d4},
the constraint~(\ref{s3rel}) implies,
given the known experimental values,
a normal mass spectrum $m_1 < m_2 < m_3$,
with $m_1$ either in the range 3 to 9$\times 10^{-3}\, \mathrm{eV}$,
or larger than 14$\times 10^{-3}\, \mathrm{eV}$;
these numbers hold for the best-fit values
of the mass-squared differences as given in~\cite{tortola}. 
A further prediction is
$\left| \left\langle m \right\rangle \right| = m_1 m_2 / m_3$,
where $\left| \left\langle m \right\rangle \right|$ 
is the effective mass
relevant for neutrinoless $\beta \beta$ decay.

\paragraph{Generalization of the $\mu$--$\tau$ interchange symmetry}

We now abandon the symmetry $\mathbbm{Z}_2^\mathrm{(e)}$
and return to the general $M_R$ of the $S_3$ model,
given in~(\ref{MR}).
Since $\mnu^{-1} = - M_D^{-1} M_R {M_D^T}^{-1}$,
and since $M_D = \mathrm{diag} \left( a, b, b \right)$ is diagonal,
one obtains $\mnu^{-1}$ of the form~(\ref{general_mnu}).

In general,
the symmetric matrix $\mnu^{-1}$ contains nine parameters:
the six moduli of its matrix
elements and three rephasing-invariant phases, 
because one may independently rephase
the three left-handed neutrinos,
thereby eliminating three phases in $\mnu$,
or equivalently in $\mnu^{-1}$.
To those nine parameters correspond nine observables:
the three neutrino masses $m_{1,2,3}$,
the three mixing angles $\theta_{12,13,23}$,
the Dirac phase $\delta$ and
the Majorana phases $\Delta$ and $\Omega$.

In the case of (full) $\mu$--$\tau$ interchange symmetry,
i.e.\ when $e^{i \psi} = \pm 1$ in~(\ref{general_mnu}),
three observables are predicted:
$\theta_{23} = \pi / 4$,
$\theta_{13} = 0$ and
the Dirac phase is meaningless because $\theta_{13} = 0$.
To those three predicted observables
correspond three rephasing-invariant relations
among the parameters of $\mnu$,
or of $\mnu^{-1}$:
\ba
\left| \left( \mnu^{-1} \right)_{e \mu} \right|
&=&
\left| \left( \mnu^{-1} \right)_{e \tau} \right|,
\label{1}
\\
\left| \left( \mnu^{-1} \right)_{\mu \mu} \right|
&=&
\left| \left( \mnu^{-1} \right)_{\tau \tau} \right|,
\label{2}
\\
\arg \left\{
\left[ \left( \mnu^{-1} \right)_{e \mu} \right]^2
\left( \mnu^{-1} \right)_{\tau \tau} \right\}
&=&
\arg \left\{
\left[ \left( \mnu^{-1} \right)_{e \tau} \right]^2
\left( \mnu^{-1} \right)_{\mu \mu} \right\}.
\label{angle_condition}
\ea

In the case of the matrix~(\ref{general_mnu}),
the condition~(\ref{angle_condition}) does not apply.
One has an \emph{incomplete $\mu$--$\tau$ interchange symmetry},
wherein conditions~(\ref{1}) and~(\ref{2}) apply,
but not condition~(\ref{angle_condition}).
The matrix~(\ref{general_mnu})
has seven real physical parameters.
As we will see,
$\psi \neq 0$ leads both $\cos 2\theta_{23}$
and $s_{13}$ to be non-zero,
and in general there will also be
a non-zero Dirac phase $\delta$.
Since the matrix~(\ref{general_mnu})
has only one parameter more than matrix~(\ref{symmetric_mnu}),
it must predict two relations:
two of the observables $\cos 2\theta_{23}$,
$s_{13}$ and $\delta$ must be functions of the third
one and of the remaining observables,
which are $m_{1,2,3}$,
$\theta_{12}$,
$\Delta$ and $\Omega$. 
Since the Majorana phases are hardly accessible by experiment, 
our aim is to derive observable consequences
of incomplete $\mu$--$\tau$ interchange symmetry
which do not involve those phases.

It is convenient to use the facts that,
from experiment,
it is known
that the atmospheric mixing angle $\theta_{23}$
is close to $\pi / 4$
and that $\theta_{13}$ is small. 
Thus we define the parameters
\ba
\nu &\equiv& \cos{2 \theta_{23}},
\\
\epsilon &\equiv& s_{13} e^{i \delta},
\ea
the latter of which is complex.
Experimentally~\cite{tortola},
$\left| \nu \right| < 0.28$ at $90\%$ confidence level
and $\left| \epsilon \right| < 0.22$ at $3 \sigma$ level.
In the case of full $\mu$--$\tau$ interchange symmetry,
$\nu = \epsilon = 0$ and there are no restrictions
on all other observables,
i.e.\ on $m_{1,2,3}$,
$\theta_{12}$,
$\Delta$ and $\Omega$.
In the case of incomplete $\mu$--$\tau$ interchange symmetry,
$\nu$ and $\epsilon$ in general do not vanish and,
when they are non-zero,
some restrictions may apply on the other observables.

Adding~(\ref{1}) and~(\ref{2}) and using~(\ref{mnu-1}),
one finds that
\ba
0 &=&
\left| \left( \mnu^{-1} \right)_{e \mu} \right|^2
+
\left| \left( \mnu^{-1} \right)_{\mu \mu} \right|^2
-
\left| \left( \mnu^{-1} \right)_{e \tau} \right|^2
-
\left| \left( \mnu^{-1} \right)_{\tau \tau} \right|^2
\\
&=&
\left( \mnu^{-1} {\mnu^{-1}}^\ast \right)_{\mu \mu}
-
\left( \mnu^{-1} {\mnu^{-1}}^\ast \right)_{\tau \tau}
\\
&=& \sum_{j=1}^3
\frac{\left| U_{\mu j} \right|^2 - \left| U_{\tau j} \right|^2}
{m_j^2}
\\ &=&
\left( c_{23}^2 - s_{23}^2 \right)
\left(
\frac{s_{12}^2 - s_{13}^2 c_{12}^2}{m_1^2}
+
\frac{c_{12}^2 - s_{13}^2 s_{12}^2}{m_2^2}
-
\frac{c_{13}^2}{m_3^2}
\right)
\no & &
+ 4 \left( \frac{1}{m_1^2} - \frac{1}{m_2^2} \right)
c_{23} s_{23} s_{13} c_{12} s_{12} \cos{\delta}.
\label{exact}
\ea
This condition is particularly useful
since it does not involve the Majorana phases.
It translates into
\be
\left(
\frac{s_{12}^2 - \left| \epsilon \right|^2 c_{12}^2}{m_1^2}
+ \frac{c_{12}^2 - \left| \epsilon \right|^2 s_{12}^2}{m_2^2}
+ \frac{\left| \epsilon \right|^2 - 1}{m_3^2}
\right) \nu
+ 2 \left( \frac{1}{m_1^2} - \frac{1}{m_2^2} \right)
 c_{12} s_{12} \sqrt{1 - \nu^2}\, \mathrm{Re}\, \epsilon
= 0.
\label{equality}
\ee
Numerically,
we shall use~(\ref{equality}) to determine $\nu$
as a function of $\epsilon$,
for various values of the neutrino masses
and of the mixing angle $\theta_{12}$.
Since $\left| \epsilon \right|^2$ and $\nu^2$
are in any case rather small,
(\ref{equality}) is an almost linear relationship
between $\nu$ and $\mathrm{Re}\, \epsilon$.

We next consider the constraint~(\ref{1}) by itself alone.
It is equivalent to the existence of a phase $\varphi$
such that
\ba
0 &=&
\left( \mnu^{-1} \right)_{e \mu}
-
e^{i \varphi} \left( \mnu^{-1} \right)_{e \tau}
\\
&=&
\sum_{j=1}^3 \frac{U_{ej}
\left( U_{\mu j} - e^{i \varphi} U_{\tau j} \right)}
{m_j}.
\label{triangle1}
\ea
This is the equation of a triangle
in the complex plane---it states that
the sum of three complex numbers vanishes,
i.e.\ that those three numbers form a triangle
in the complex plane.
The triangle~(\ref{triangle1}) involves the Majorana phases.
It is convenient to remove those phases,
since they are in practice
very difficult to observe experimentally.
One does this by considering the inequality,
which follows from~(\ref{triangle1}),\footnote{It is possible
to construct a triangle with sides of
(real, non-negative)
lengths $a$,
$b$ and $c$ if and only if
\[
a\, \le\, b+c, \quad b\, \le\, a+c
\quad \mathrm{and} \quad c\, \le\, a+b.
\]
It is easily shown that this set of three inequalities
is equivalent to the sole inequality~\cite{book}
\[
a^4 + b^4 + c^4 - 2 \left( a^2 b^2 + a^2 c^2 + b^2 c^2 \right)
\le\, 0.
\]
}
\be
\sum_{j=1}^3 \frac{\left| U_{ej}
\left( U_{\mu j} - e^{i \varphi} U_{\tau j} \right) \right|^4}
{m_j^4}
- 2 \sum_{j < k}
\frac{\left| U_{ej}
\left( U_{\mu j} - e^{i \varphi} U_{\tau j} \right) \right|^2
\left| U_{ek}
\left( U_{\mu k} - e^{i \varphi} U_{\tau k} \right) \right|^2}
{m_j^2 m_k^2}\ \le\, 0.
\label{triangle2}
\ee
Notice that,
using~(\ref{formU}),
one has
\ba
\left| U_{e1} \left( U_{\mu 1}
- e^{i \varphi} U_{\tau 1} \right) 
\right|^2
\left/ c_{13}^2 \right. &=&
c_{12}^2 s_{12}^2 \left( 1
- \sqrt{1 - \nu^2} \cos{\phi} \right)
\no & &
+ c_{12}^4 \left| \epsilon \right|^2 \left( 1
+ \sqrt{1 - \nu^2} \cos{\phi} \right)
\no & &
+ 2 c_{12}^3 s_{12} \left( \nu\, \mathrm{Re}\, \epsilon \cos{\phi}
- \mathrm{Im}\, \epsilon \sin{\phi} \right),
\label{eq1} \\
\left| U_{e2} \left( U_{\mu 2}
- e^{i \varphi} U_{\tau 2} \right) 
\right|^2
\left/ c_{13}^2 \right. &=&
c_{12}^2 s_{12}^2 \left( 1
- \sqrt{1 - \nu^2} \cos{\phi} \right)
\no & &
+ s_{12}^4 \left| \epsilon \right|^2 \left( 1
+ \sqrt{1 - \nu^2} \cos{\phi} \right)
\no & &
- 2 c_{12} s_{12}^3 \left( \nu\, \mathrm{Re}\, \epsilon \cos{\phi}
- \mathrm{Im}\, \epsilon \sin{\phi} \right),
\label{eq2} \\
\left| U_{e3} \left( U_{\mu 3}
- e^{i \varphi} U_{\tau 3} \right) 
\right|^2
\left/ c_{13}^2 \right. &=&
\left| \epsilon \right|^2 \left( 1
+ \sqrt{1 - \nu^2} \cos{\phi} \right),
\label{eq3}
\ea
where $\phi \equiv \varphi + \vartheta_3 - \vartheta_2$.

Numerically,
we use~(\ref{equality}) to determine $\nu$
as a function of $\epsilon$,
for various values of the neutrino masses
and of the mixing angle $\theta_{12}$.
Afterwards,
we check whether there is any phase $\varphi$
for which the inequality~(\ref{triangle2}) is satisfied.
If there is,
then those values of the neutrino masses,
mixing angles and Dirac phase are compatible with
incomplete $\mu$--$\tau$ interchange symmetry;
otherwise they are not.
For simplicity
we keep $\theta_{12} = 33^\circ$,
$m_2^2 - m_1^2 = 8.1 \times 10^{-5}\, \mathrm{eV}^2$
and $\left| m_3^2 - m_1^2 \right| = 2.2 \times 10^{-3}\, \mathrm{eV}^2$
fixed at their best-fit values~\cite{tortola}.
It is important to remark that~(\ref{eq1})--(\ref{eq3}),
just as~(\ref{equality}),
are symmetric under $\nu \to - \nu,\
\mathrm{Re}\, \epsilon \to - \mathrm{Re}\, \epsilon$.
This means that one only has to study the region
of positive $\mathrm{Re}\, \epsilon$.
Also,
(\ref{eq1})--(\ref{eq3}) are invariant under
$\mathrm{Im}\, \epsilon \to - \mathrm{Im}\, \epsilon,\
\sin{\phi} \to - \sin{\phi}$.
This means that we only have to consider positive values
of $\mathrm{Im}\, \epsilon$,
provided we test all possible values of $\phi$.

\begin{figure}[htb]
\centering
\includegraphics[clip,height=80mm]{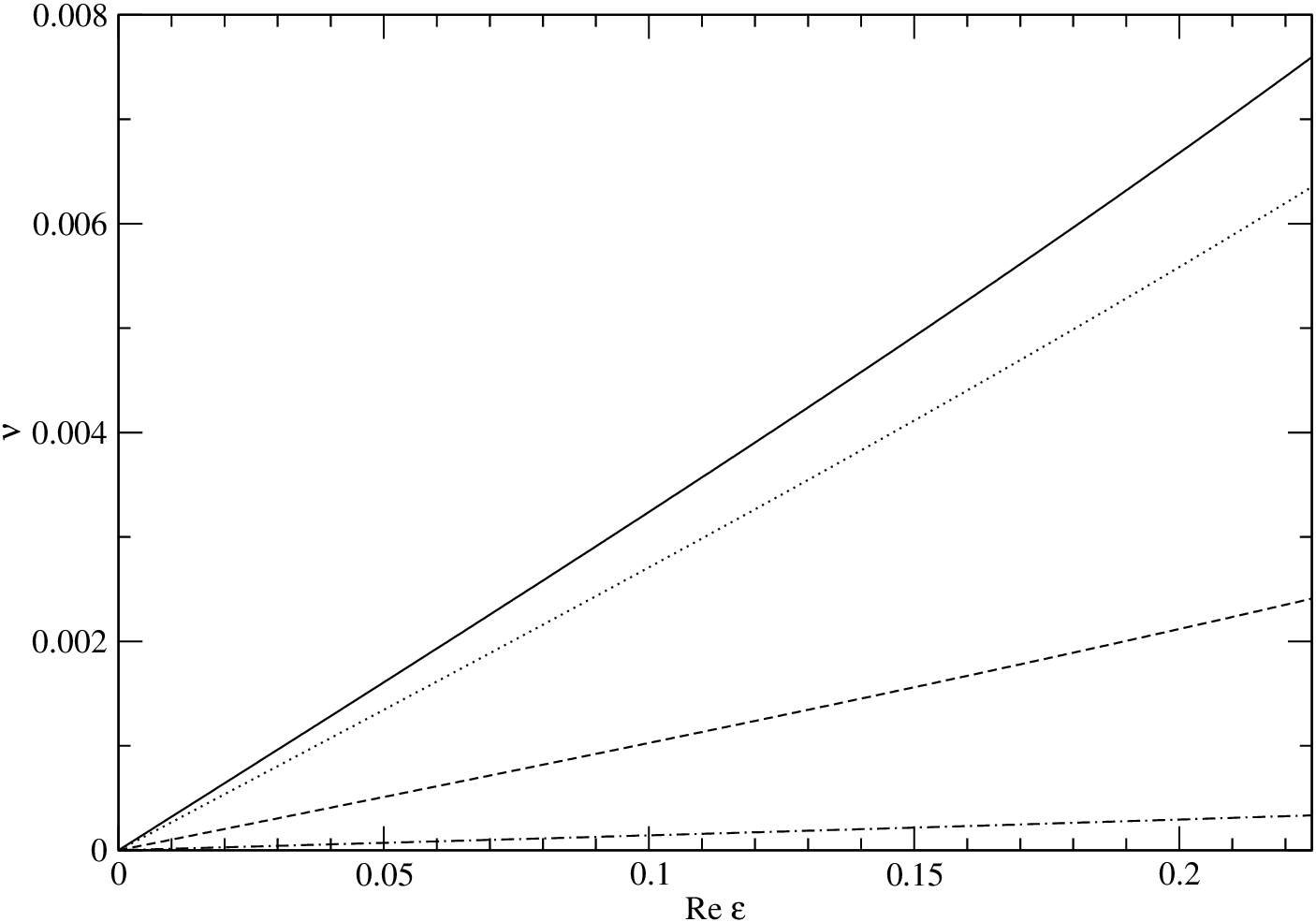}
\caption{$\nu$ as a function of $\mathrm{Re}\, \epsilon$
in the case $m_3 < m_{1,2}$.
The figure corresponds to a vanishing $\mathrm{Im}\, \epsilon$,
but it would be practically identical
for any $\mathrm{Im}\, \epsilon$
of order $0.1$ or smaller.
The full line is for $m_3^2 = 10^{-1}\, \mathrm{eV}^2$,
the dotted line for $m_3^2 = 10^{-2}\, \mathrm{eV}^2$,
the dashed line for $m_3^2 = 10^{-3}\, \mathrm{eV}^2$
and the dashed-dotted line for $m_3^2 = 10^{-4}\, \mathrm{eV}^2$.}
\label{fig1}
\end{figure}

In the case where $m_3^2 - m_1^2 < 0$,
the situation is rather simple
and it is aptly described
by fig.~\ref{fig1}.
The parameter $\nu$ has the same sign as $\mathrm{Re}\, \epsilon$
but it is much smaller in absolute value;
the atmospheric mixing angle is,
for all practical purposes,
maximal;
in the limit of very small $m_3$
the relation $\theta_{23} = 45^\circ$
becomes exact---see~(\ref{equality}) and fig.~\ref{fig1}.
The exact value of $\mathrm{Im}\, \epsilon$
is practically immaterial in the determination of $\nu$
as a function of $\mathrm{Re}\, \epsilon$.
The inequality~(\ref{triangle2}) is always satisfied,
hence it has no bearing on the overall picture.

\begin{figure}[t]
\centering
\includegraphics[clip,height=80mm]{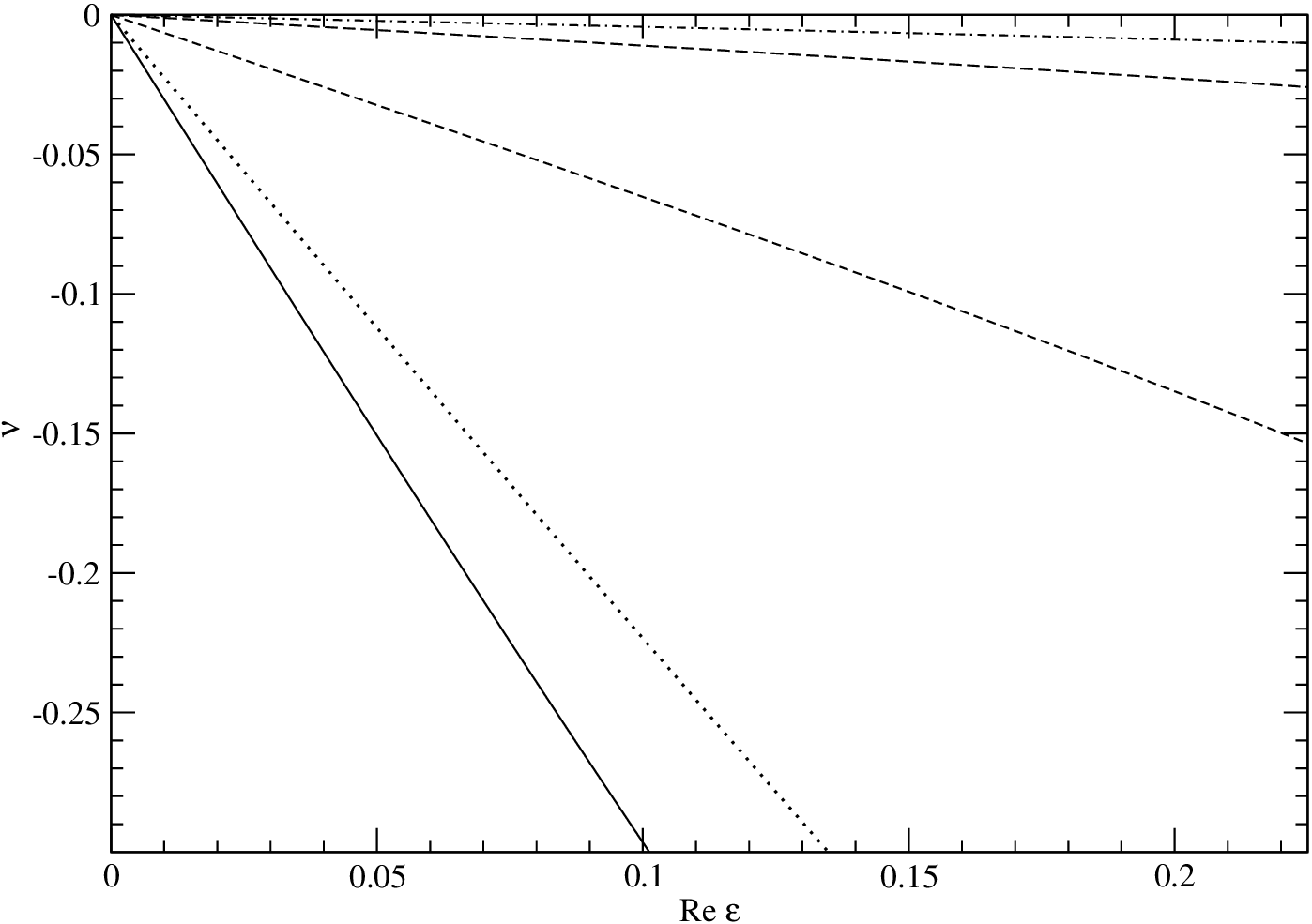}
\caption{$\nu$ as a function of $\mathrm{Re}\, \epsilon$
in the case $m_3 > m_{1,2}$
and with $\mathrm{Im}\, \epsilon = 0.1$
(a smaller $\mathrm{Im}\, \epsilon$
yields practically the same curves,
except for the exclusion zones depicted in fig.~\ref{fig3}).
The full line is for $m_1^2 = 10^{-6}\, \mathrm{eV}^2$,
the dotted line for $m_1^2 = 10^{-5}\, \mathrm{eV}^2$,
the small-dashed line for $m_1^2 = 10^{-4}\, \mathrm{eV}^2$,
the large-dashed line for $m_1^2 = 10^{-3}\, \mathrm{eV}^2$
and the dashed-dotted line for $m_1^2 = 10^{-2}\, \mathrm{eV}^2$.}
\label{fig2}
\end{figure}

In the case where $m_3^2 - m_1^2 > 0$
the situation is different.
The parameters $\nu$ and $\mathrm{Re}\, \epsilon$
have opposite signs
and $\nu$ is not necessarily small.
On the other hand,
the determination of $\nu$ as a function
of $\mathrm{Re}\, \epsilon$ is,
once again,
largely insensitive to the exact value
of $\mathrm{Im}\, \epsilon$.
Typical values are displayed in fig.~\ref{fig2}.

\begin{figure}[b]
\centering
\includegraphics[clip,height=80mm]{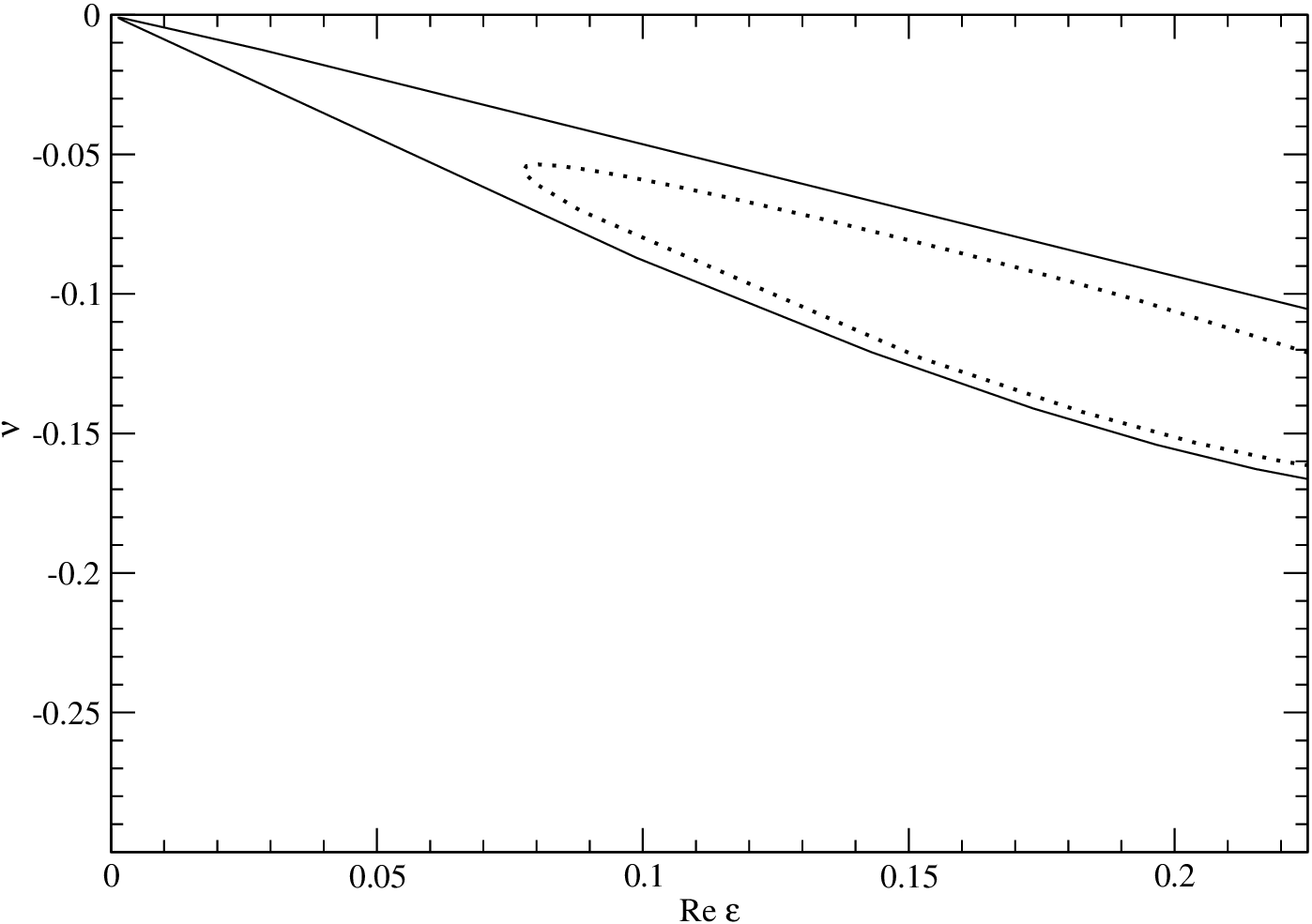}
\caption{In the case $m_3 > m_{1,2}$,
the region in the $\nu$--$\mathrm{Re}\, \epsilon$ plane
inside the full line is excluded by~(\ref{triangle2})
when $\mathrm{Im}\, \epsilon = 10^{-4}$.
The region inside the dotted line is excluded
when $\mathrm{Im}\, \epsilon = 5 \times 10^{-3}$.
For $\mathrm{Im}\, \epsilon$ larger than $10^{-2}$
there is no excluded region any more.}
\label{fig3}
\end{figure}

When $m_3^2 - m_1^2 > 0$, 
inequality~(\ref{triangle2}) introduces a complication 
because in this case
there are values of the pair $\left( \nu, \epsilon \right)$ 
for which that inequality is satisfied
by no phase $\varphi$ at all.
With $\nu$ and $\mathrm{Re}\, \epsilon$
obeying the relation~(\ref{equality}),
this happens when $\mathrm{Im}\, \epsilon \lesssim 0.01$
and $m_1 \sim 10^{-2}\, \mathrm{eV}$.
For $\mathrm{Im}\, \epsilon = 10^{-4}$ and $5 \times 10^{-3}$,
the corresponding excluded regions
in the $\left( \nu, \mathrm{Re}\, \epsilon \right)$ plane 
are depicted in fig.~\ref{fig3}.
That figure should be superimposed on fig.~\ref{fig2}
in order to see which curves
or which parts of the curves
in that figure are excluded
and to find out the range of values of $m_1^2$ 
for which an excluded region arises.\footnote{In this comparison
we use the fact that the curves in fig.~\ref{fig2}
are practically independent
of the value of $\mathrm{Im}\, \epsilon$,
provided $\mathrm{Im}\, \epsilon \lesssim 0.1$.}
Of the curves depicted in fig.~\ref{fig2},
only the small-dashed line,
referring to $m_1 = 10^{-2}\, \mathrm{eV}$,
is affected:
for $\mathrm{Im}\, \epsilon = 10^{-4}$
that line is almost completely excluded,
whereas for $\mathrm{Im}\, \epsilon = 5 \times 10^{-3}$
it is excluded if $\mathrm{Re}\, \epsilon \gtrsim 0.08$.
Explicitly,
we have found that,
when $\mathrm{Im}\, \epsilon$ vanishes,
excluded values of $\nu$ and $\mathrm{Re}\, \epsilon$ arise for
$7.80 \times 10^{-3}\, \mathrm{eV}
< m_1 < 1.28 \times 10^{-2}\, \mathrm{eV}$;
when $\mathrm{Im}\, \epsilon = 5 \times 10^{-3}$,
excluded values of $\mathrm{Re}\, \epsilon$ arise only
for $m_1$ in between $8.49 \times 10^{-3}\, \mathrm{eV}$
and $1.16 \times 10^{-2}\, \mathrm{eV}$.

The excluded regions in the
$\left( \nu, \mathrm{Re}\, \epsilon \right)$ plane
can be translated into lower bounds
for the $CP$-violating phase $\delta$.
These lower bounds are functions of $m_1$ and of $s_{13}$.
Taking $m_1 = 0.01\, \mathrm{eV}$,
i.e.\ $m_1$ in the center of the range
where excluded regions occur,
we numerically find
$\delta \gtrsim 2.44^\circ$ for $s_{13} = 0.2$, 
$\delta \gtrsim 3.34^\circ$ for $s_{13} = 0.1$,
$\delta \gtrsim 3.52^\circ$ for $s_{13} = 0.01$,
and $\delta \gtrsim 3.83^\circ$ for $s_{13} = 0.001$. 
(We have confined ourselves
to $\delta$ in the first quadrant,
i.e.\ to the real and imaginary part of $\epsilon$ being both positive.
The bounds on $\delta$ in the first quadrant
get transferred into the other quadrants
by using the symmetries
$\mathrm{Re}\, \epsilon \to - \mathrm{Re}\, \epsilon,\
\nu \to - \nu$
and $\mathrm{Im}\, \epsilon \to - \mathrm{Im}\, \epsilon,\
\sin{\phi} \to - \sin{\phi}$ referred to earlier.)
One sees that the excluded domain is hardly significant
in terms of $\delta$.
For $s_{13} \gtrsim 0.1$,
this is qualitatively understandable from the fact that
for $\mathrm{Im}\, \epsilon = s_{13} \sin \delta \gtrsim 0.01$ 
there is no exclusion region anymore.

\paragraph{Conclusions}
In this paper
we have considered an extension of the Standard Model
based on the horizontal symmetry group $S_3 \times \mathbbm{Z}_2$, 
the seesaw mechanism and a complex scalar gauge singlet $\chi$.
Though $S_3$ is a time-honoured symmetry,
the new feature here is
the use of the complex scalar gauge singlet,
with $(\chi, \chi^*)$ transforming
as a two-dimensional irreducible
representation of $S_3$---see~(\ref{Z2tr}) and (\ref{z3}).
The gauge multiplets of our extension
are those of the Standard Model,
supplemented by two additional Higgs doublets,
the scalar singlet
and three right-handed neutrino singlets for the seesaw mechanism.
The horizontal symmetry enforces diagonal
charged-lepton and neutrino Dirac mass matrices.
Our model has some freedom
with regard to the realization of the symmetry
$S_3 \times \mathbbm{Z}_2$,
and this freedom affects the Majorana mass matrix
of the right-handed neutrinos.
In this way,
we are able to recover two mass matrices
already found in the literature,
derived from different horizontal symmetries,
and also the mass matrix~(\ref{general_mnu}),
a generalization thereof;
we regard this flexibility as the  
distinguishing feature of the $S_3 \times \mathbbm{Z}_2$ model.
In terms of the
inverted light-neutrino mass matrix~(\ref{general_mnu}),
our results can be described in the following way: 
\begin{enumerate}
\item Imposing the additional discrete electron number 
$\mathbbm{Z}_2^{(e)}$ of~(\ref{Z2e}),
one obtains $\psi = 0$ and $q = 0$;
one thus recovers a mass matrix
originally derived in~\cite{d4}
from a horizontal symmetry group $D_4$.
\item Without $\mathbbm{Z}_2^{(e)}$ and with exact
$S_3 \times \mathbbm{Z}_2$ symmetry of the Lagrangian,
one gets $\psi = 0$,
i.e.\ the $\mu$--$\tau$ symmetric mass matrix
originally obtained in~\cite{z2}
in a framework of softly broken lepton numbers. 
\item Breaking $S_3 \times \mathbbm{Z}_2$
softly in the scalar potential,
one obtains the matrix~(\ref{general_mnu})
without further restrictions.
\end{enumerate}
The first and second realizations
have a $\mu$--$\tau$ symmetric mass matrix,
with the well-known predictions of
maximal atmospheric neutrino mixing
and vanishing mixing-matrix element $U_{e3}$.
If,
in addition,
$q=0$ holds,
then the model becomes much more predictive: 
it requires a normal neutrino mass ordering $m_1 < m_2 <m_3$, 
and the effective mass
in neutrinoless $\beta \beta$ decay
is a simple function of the neutrino masses alone~\cite{d4}.
Below the seesaw scale,
the first and second realizations
cannot be distinguished from the models in~\cite{d4} and~\cite{z2},
respectively. 

With $\psi \neq 0$,
the mass matrix~(\ref{general_mnu})
has seven physical parameters
and partially breaks the $\mu$--$\tau$ interchange symmetry;
the matrix of the absolute values
of the elements of the \emph{inverted}
neutrino mass matrix~(\ref{general_mnu})
is still $\mu$--$\tau$ 
symmetric.\footnote{Any mass matrix with that property
can be transformed into~(\ref{general_mnu}) by a phase transformation.}
In contrast to full $\mu$--$\tau$ interchange symmetry, 
this partial symmetry induces non-zero 
$\cos{2 \theta_{23}}$ and $s_{13}$,
and $CP$ violation in neutrino mixing via the Dirac phase $\delta$.
These three quantities are functions of $\psi$.
Fixing the neutrino masses
and the solar mixing angle,
there is an almost linear relation~(\ref{equality})
between $\cos{2 \theta_{23}}$
and $s_{13} \cos\delta$,
which is not obfuscated by the Majorana phases.
This relation differs substantially
depending on the type of neutrino mass spectrum:
in the inverted case,
atmospheric mixing is always maximal for all practical purposes,
even when $s_{13}$ is close to its experimental upper bound;
in the normal case,
a large $s_{13} \cos\delta$ is correlated
with a large $\cos{2 \theta_{23}}$ with an opposite sign.

Finally,
as an additional virtue,
we mention that,
for a normal neutrino mass spectrum, 
leptogenesis can naturally be accommodated in the present model
with the $\mu$--$\tau$ symmetric
mass matrices~\cite{leptogenesis};
at least for small $s_{13}$,
the same must hold with partial $\mu$--$\tau$ interchange symmetry.

\vspace*{5mm}

\paragraph{Acknowledgement}
The work of L.L.\ was supported by the Portuguese
\textit{Funda\c c\~ao para a Ci\^encia e a Tecnologia}
through the projects POCTI/FNU/44409/2002 
and U777--Plurianual.

\newpage

\newpage


\end{document}